\documentclass[12pt]{article}

\usepackage{epsfig}

\newlength{\dinwidth} \newlength{\dinmargin}
\begin{document}

\begin{center}
{\Large \bf Top quark cross sections and differential distributions}
\end{center}
\vspace{1mm}
\begin{center}
{\large Nikolaos Kidonakis}\\
\vspace{2mm}
{\it Kennesaw State University,  Physics \#1202, Kennesaw, GA 30144, USA}
\end{center}

\begin{abstract}
I present results for the top quark pair total cross section
and the top quark transverse momentum distribution at Tevatron and LHC energies.
I also present results for single top quark production. All calculations
include NNLO corrections from NNLL threshold resummation.
\end{abstract}

\vspace{3mm}

\begin{center}
{\large \bf TOP-ANTITOP PAIR PRODUCTION}
\end{center}

The leading order processes for top-antitop pair production
are $q{\bar q} \rightarrow t {\bar t}$ (dominant at the Tevatron) 
and $gg \rightarrow t {\bar t}$ (dominant at LHC energies).
The QCD corrections for top pair production are significant  
and receive contributions from soft-gluon corrections which are dominant near 
threshold. These soft corrections have been resummed through 
NNLL \cite{NKttbar}, requiring two-loop calculations of the soft 
anomalous dimensions \cite{NKttbar,NK2l}. 
Approximate NNLO differential-level cross sections, using  
single-particle inclusive kinematics for partonic threshold, 
can be derived from the expansion of the resummed cross section.

Figures 1 and 2 show the NNLO approximate cross section \cite{NKttbar} together with recent data from the corresponding experiments at the Tevatron 
\cite{CDFtt,D0tt} and the LHC \cite{ATLAStt,CMStt}. The theoretical 
prediction agrees well with the measured cross sections. The upper 
and lower curves indicate the uncertainty from scale variation and pdf errors.
It is important to note that the soft-gluon approximation works very well 
not only for Tevatron but also for LHC 
energies because partonic threshold is still important.
There is only 1\% difference between the first-order approximate and exact 
corrections as shown in Fig. 3, 
and thus less than 1\% difference between NLO approximate and exact cross 
sections. For our best prediction in Figs. 1 and 2 we added the NNLO approximate 
corrections to the exact NLO cross section. In all the results presented here 
we have used the MSTW 2008 NNLO pdf \cite{MSTW08}. 

At the Tevatron, we find that the NNLO corrections 
provide a 7.8\% enhancement over NLO. For a top quark mass of 173 GeV, we find 
$$
\sigma^{\rm NNLOapprox}_{t{\bar t}}(m_t=173 \, {\rm GeV}, \, 1.96\, 
{\rm TeV})=7.08 {}^{+0.00}_{-0.24} {}^{+0.36}_{-0.27} \; {\rm pb} 
$$
where the first uncertainty is from scale variation between $m_t/2$ and $2m_t$ 
and the second is from the MSTW NNLO pdf at 90\% C.L.
The NNLO approximate corrections reduce the scale dependence greatly 
over a large range; the separate factorization and renormalization scale 
dependence has also been calculated in \cite{NKttbar}.

\begin{figure}
\begin{center}
  \includegraphics[width=11cm]{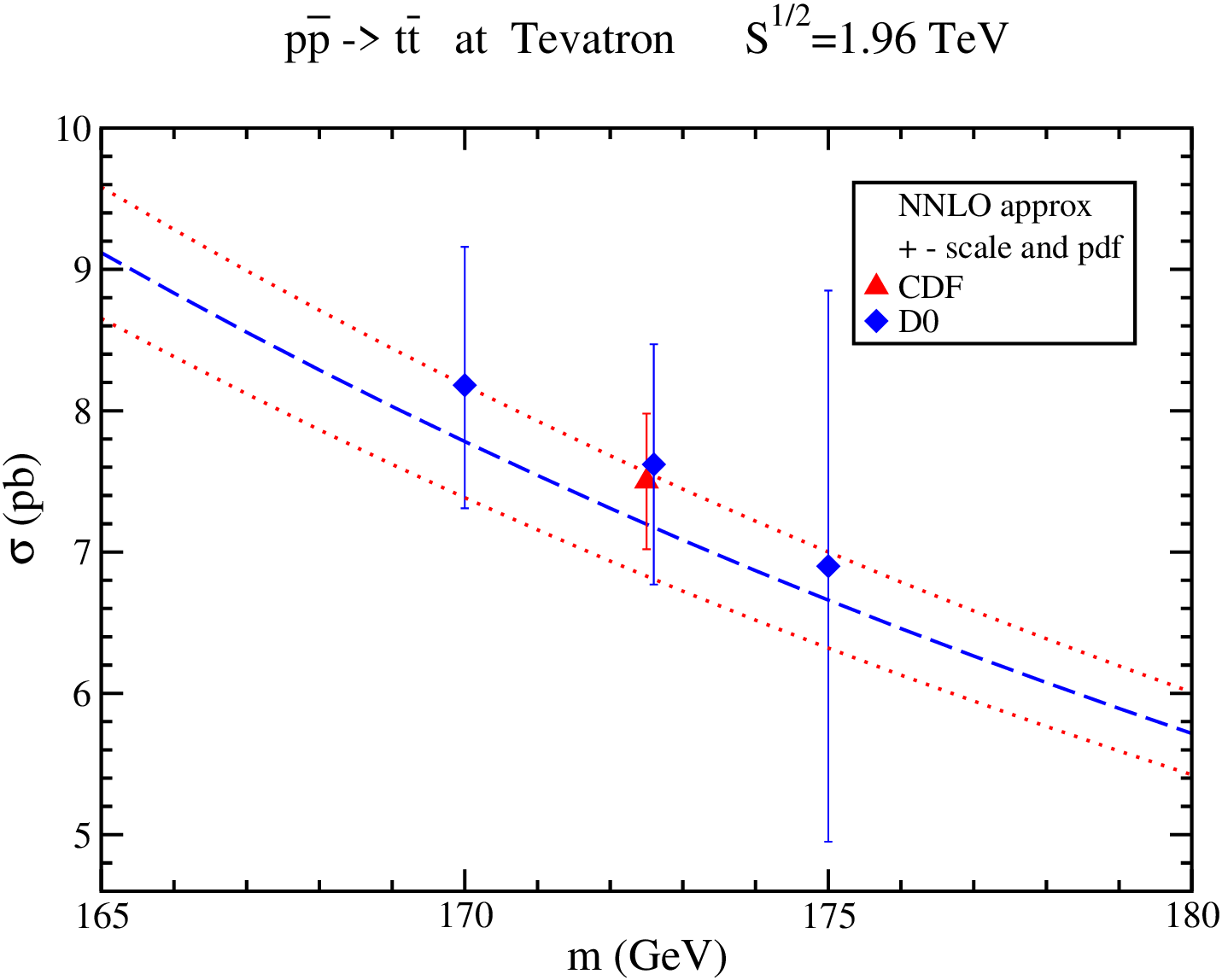}
  \caption{Top-antitop pair cross section at the Tevatron.}
\end{center}
\end{figure}

At the LHC at 7 TeV energy, we find
$$
\sigma^{\rm NNLOapprox}_{t{\bar t}}(m_t=173\, {\rm GeV}, \, 7\, 
{\rm TeV})=163 {}^{+7}_{-5}  {}^{+9}_{-9} \; {\rm pb},
$$
which is an enhancement over NLO of 7.6\%. 

\begin{figure}
\begin{center}
  \includegraphics[width=11cm]{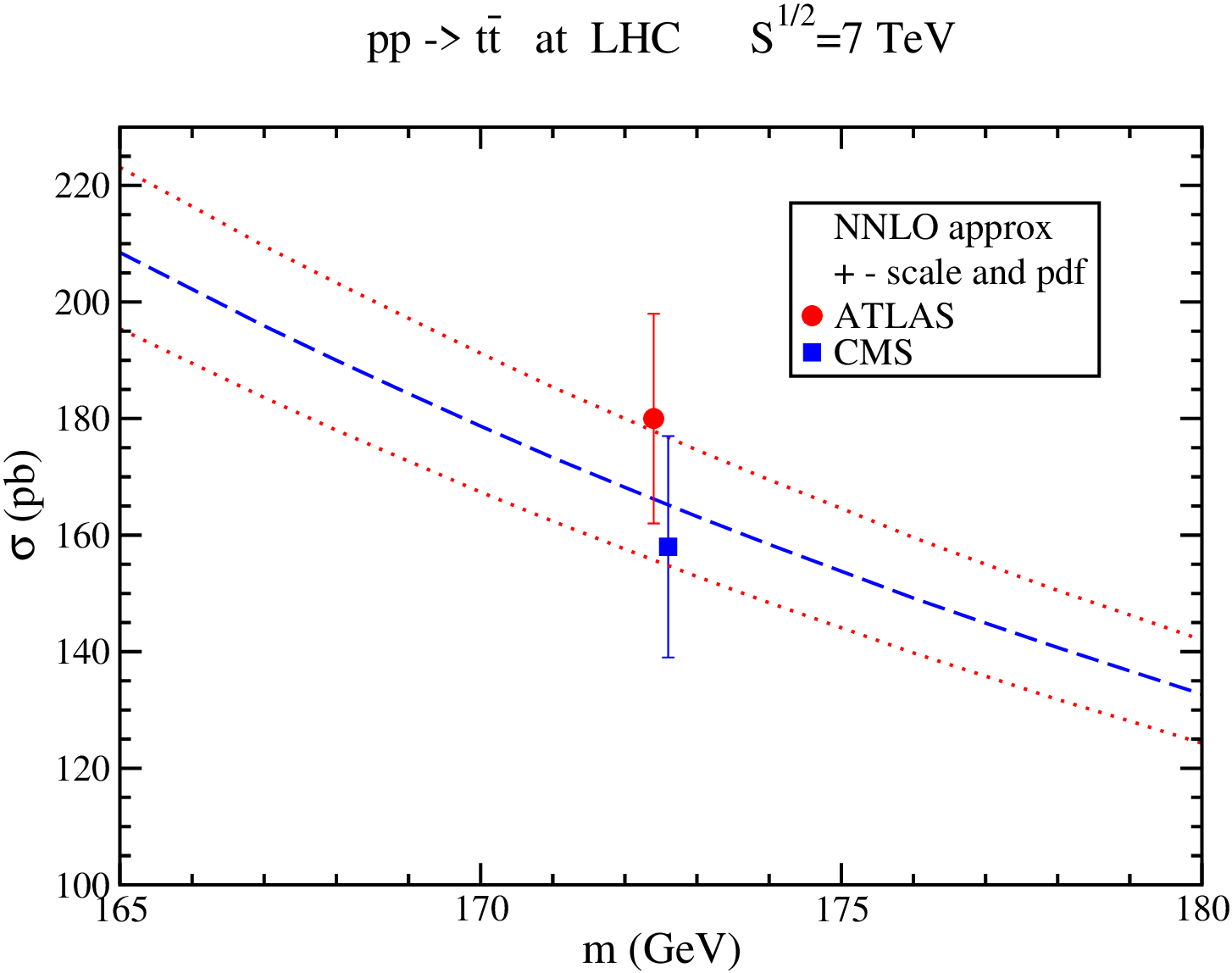}
  \caption{Top-antitop pair cross section at the LHC.}
\end{center}
\end{figure}

\begin{figure}
\begin{center}
  \includegraphics[width=11cm]{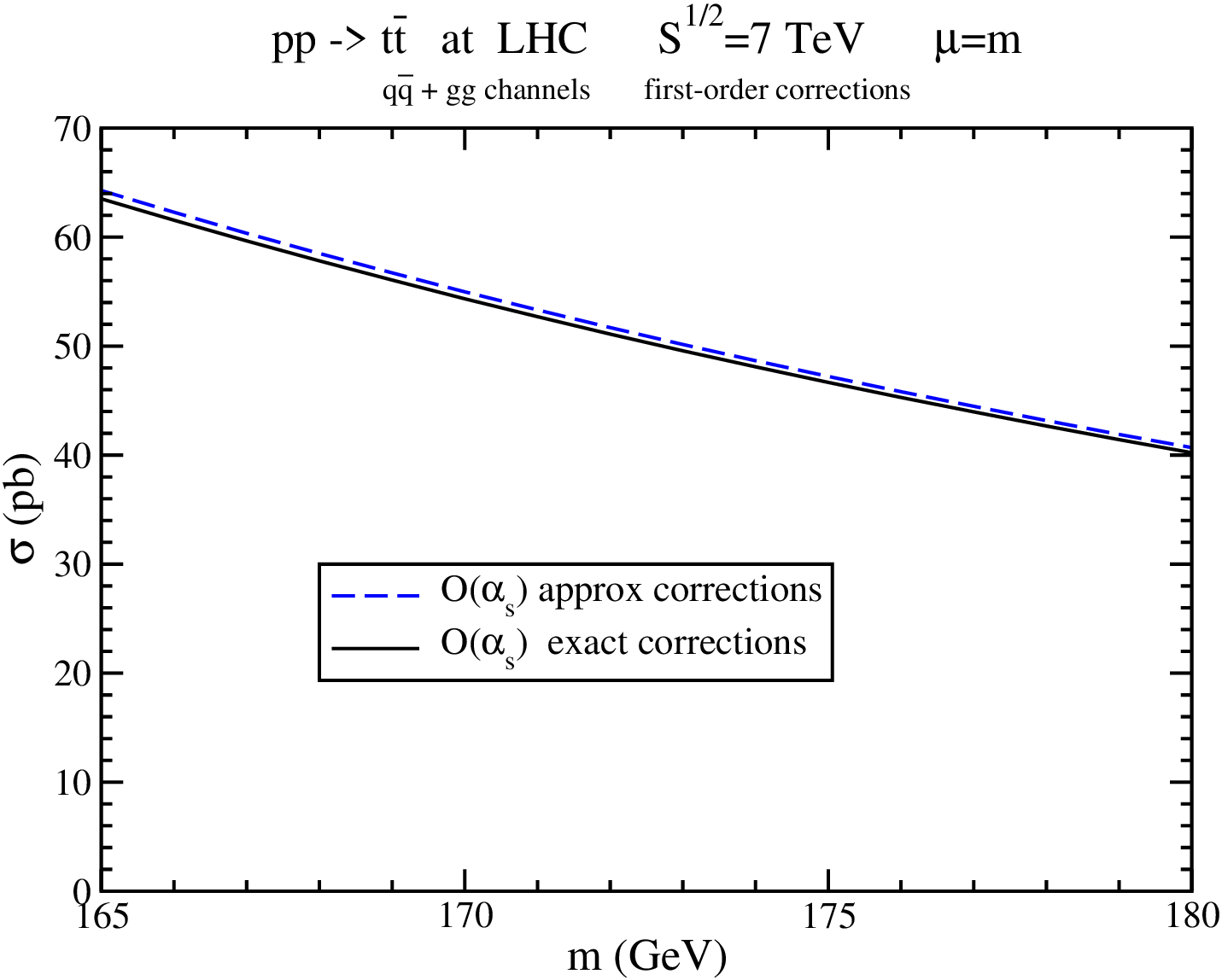}
  \caption{Approximate and exact NLO corrections for $t{\bar t}$ production 
at the LHC.}
\end{center}
\end{figure}

The top quark transverse momentum distribution at the Tevatron is shown 
in Fig. 4. The $p_T$ distribution is enhanced by the 
NNLO corrections but the shape is not significantly affected. Similar 
results have also been obtained for the LHC \cite{NKttbar}.

\begin{figure}
\begin{center}
  \includegraphics[width=11cm]{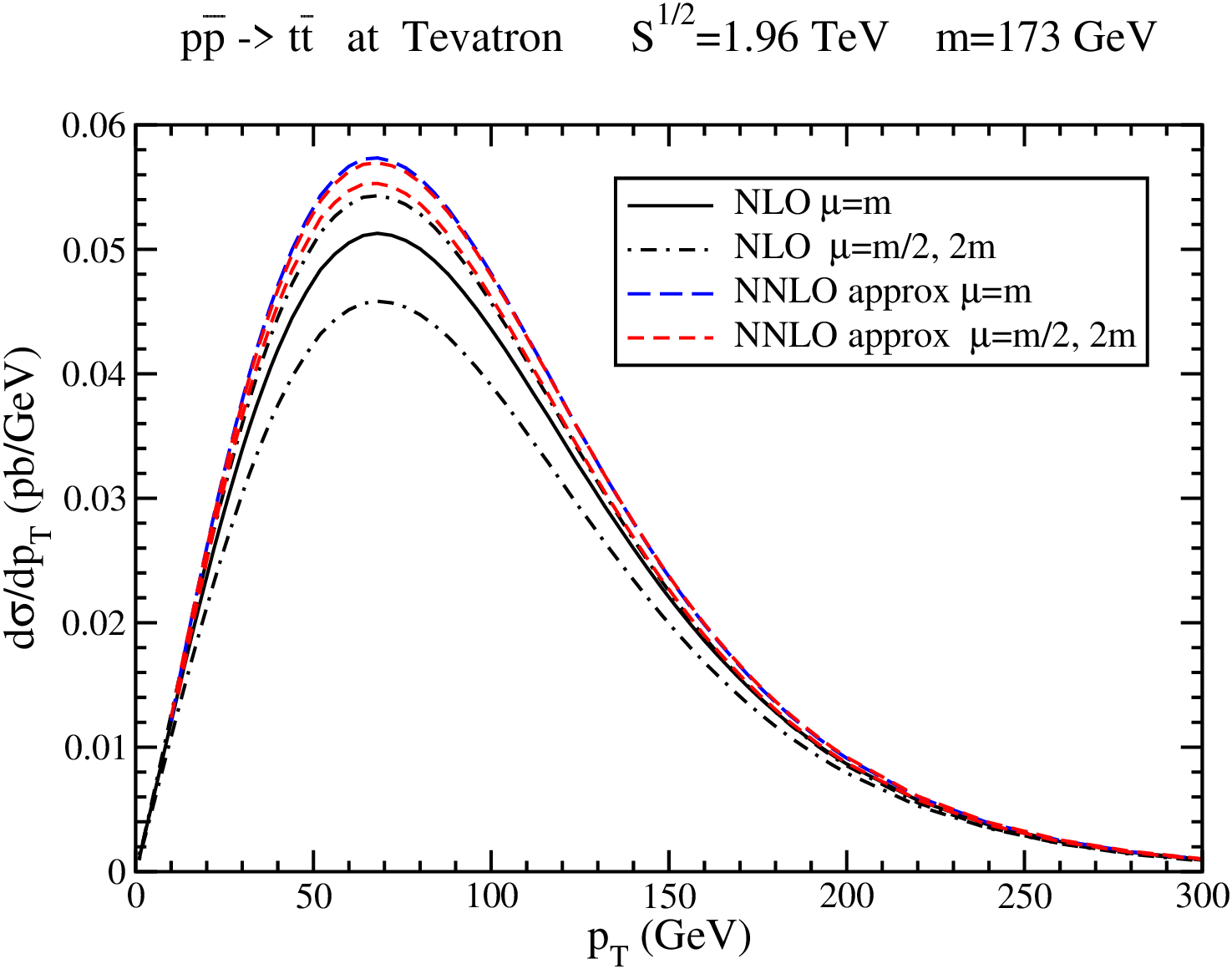}
  \caption{Top quark $p_T$ distribution at the Tevatron.}
\end{center}
\end{figure}

In Fig. 5 we show the theoretical cross sections for 
$t{\bar t}$ production in $p{\bar p}$ and $pp$ collisions as functions of 
collider energy and corresponding Tevatron \cite{CDFtt,D0tt} 
and LHC \cite{ATLAStt,CMStt} data, again noting the 
agreement between theory and experiment.

\begin{figure}
\begin{center}
  \includegraphics[width=11cm]{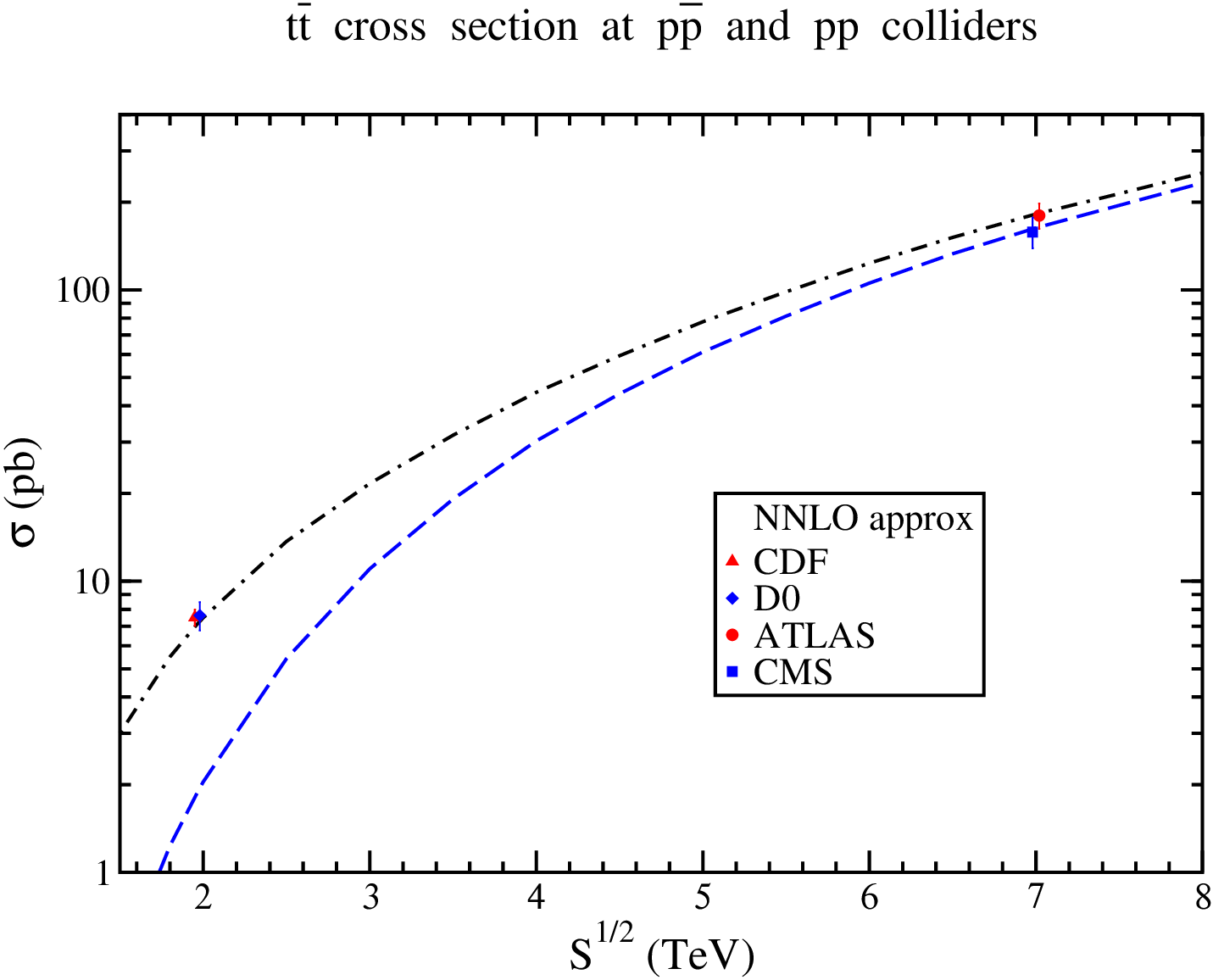}
  \caption{Top-antitop pair cross sections in $p{\bar p}$ (dash-dotted line) 
and $pp$ (dashed line) collisions versus collider energy.}
\end{center}
\end{figure}

\vspace{3mm}

\begin{center}
{\large \bf SINGLE TOP QUARK PRODUCTION}
\end{center}

We continue with single top quark production and start by discussing  
the $t$-channel processes: $qb \rightarrow q' t$ and ${\bar q} b 
\rightarrow {\bar q}' t$. The $t$ channel is numerically the largest 
at the Tevatron and the LHC.
We find for the NNLO approximate cross section \cite{NKtch} 
$$
\sigma^{\rm NNLOapprox,\, top}_{t-{\rm channel}}(m_t=173 \, {\rm GeV
}, \, 1.96\, {\rm TeV})=1.04 {}^{+0.00}_{-0.02} \pm 0.06 \; {\rm pb},
$$
$$
\sigma^{\rm NNLOapprox,\, top}_{t-{\rm channel}}(m_t=173 \, {\rm GeV
}, \, 7\, {\rm TeV})=41.7 {}^{+1.6}_{-0.2} \pm 0.8 \; {\rm pb}.
$$
The NNLO approximate corrections contribute a 4\% increase over NLO at the Tevatron and 
a 1\% decrease at the LHC at 7 TeV.

For $t$-channel antitop production the cross section at the Tevatron is 
identical to that for top production. However, at the LHC 
the cross section is different
$$
\sigma^{\rm NNLOapprox,\, antitop}_{t-{\rm channel}}(m_t=173 \, {\rm
 GeV}, \, 7\, {\rm TeV})=22.5 \pm 0.5 {}^{+0.7}_{-0.9} \; {\rm pb}.
$$

Figure 6 shows the combined single top plus single antitop 
$t$-channel cross section as a function of energy together with data from the 
Tevatron \cite{D0tch,CDFtch} and the LHC \cite{CMStch,ATLAStch}. Again, the theory 
is consistent with the measured cross sections.

\begin{figure}
\begin{center}
  \includegraphics[width=11cm]{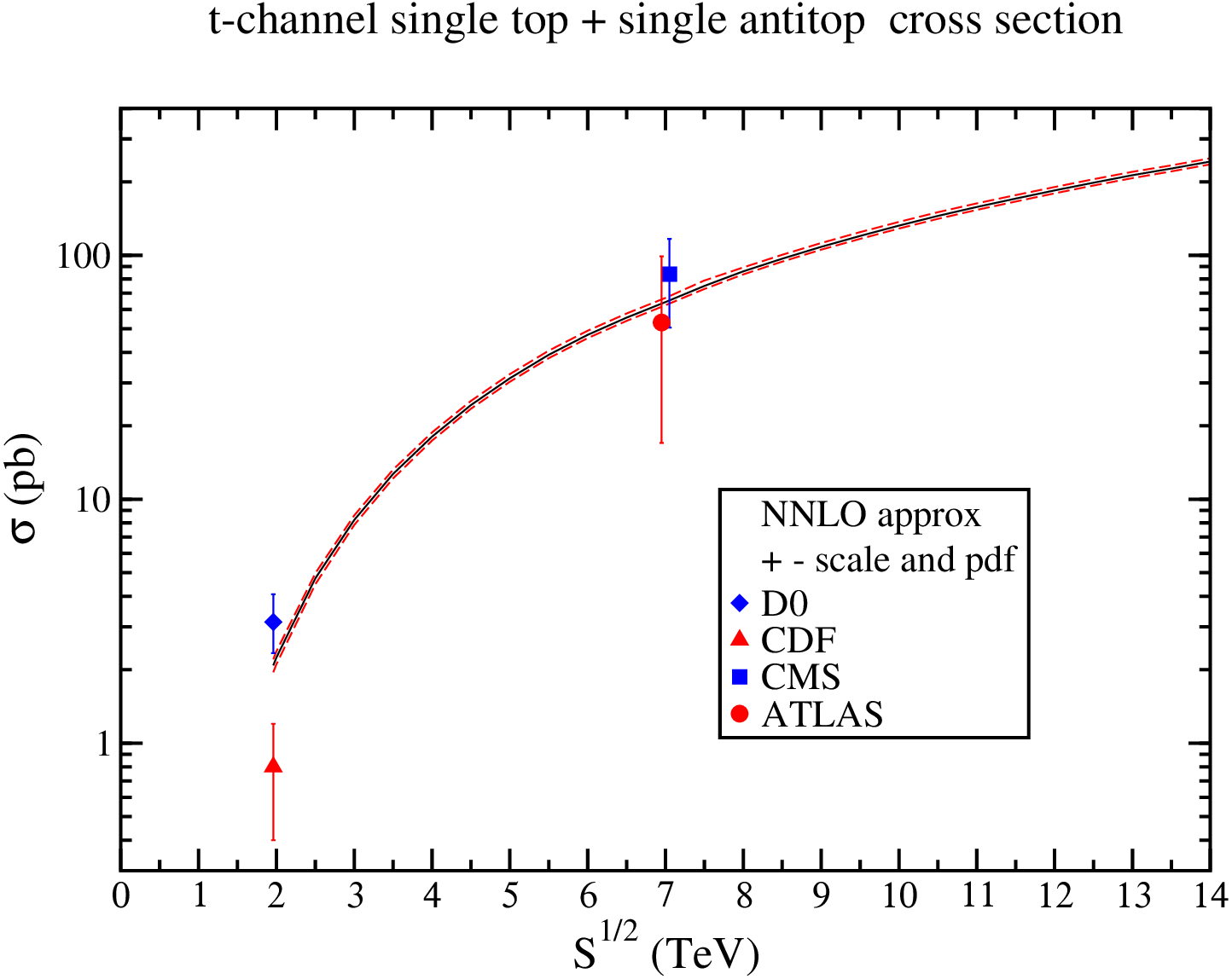}
  \caption{Single top plus single antitop $t$-channel cross section versus collider energy.}
\end{center}
\end{figure}

We continue with $s$-channel single top quark production: $q{\bar q}' \rightarrow {\bar b} t$,
which is numerically small at both Tevatron and LHC energies \cite{NKsch}.
For top production we find
$$
\sigma^{\rm NNLOapprox,\, top}_{s-{\rm channel}}(m_t=173 \, {\rm GeV
}, \, 1.96\, {\rm TeV})=0.523 {}^{+0.001}_{-0.005} {}^{+0.030}_{-0.028} \; {\rm pb},
$$
$$
\sigma^{\rm NNLOapprox,\, top}_{s-{\rm channel}}(m_t=173\, {\rm GeV
}, \, 7\, {\rm TeV})=3.17 \pm 0.06 {}^{+0.13}_{-0.10} \; {\rm pb}.
$$
The NNLO approximate corrections are an enhancement over NLO of 15\% at the 
Tevatron and 13\% at the LHC.
Figure 7 shows the $s$-channel top cross section as a 
function of top quark mass at the Tevatron. The antitop cross section 
at the Tevatron is the same.  

\begin{figure}
\begin{center}
  \includegraphics[width=10cm]{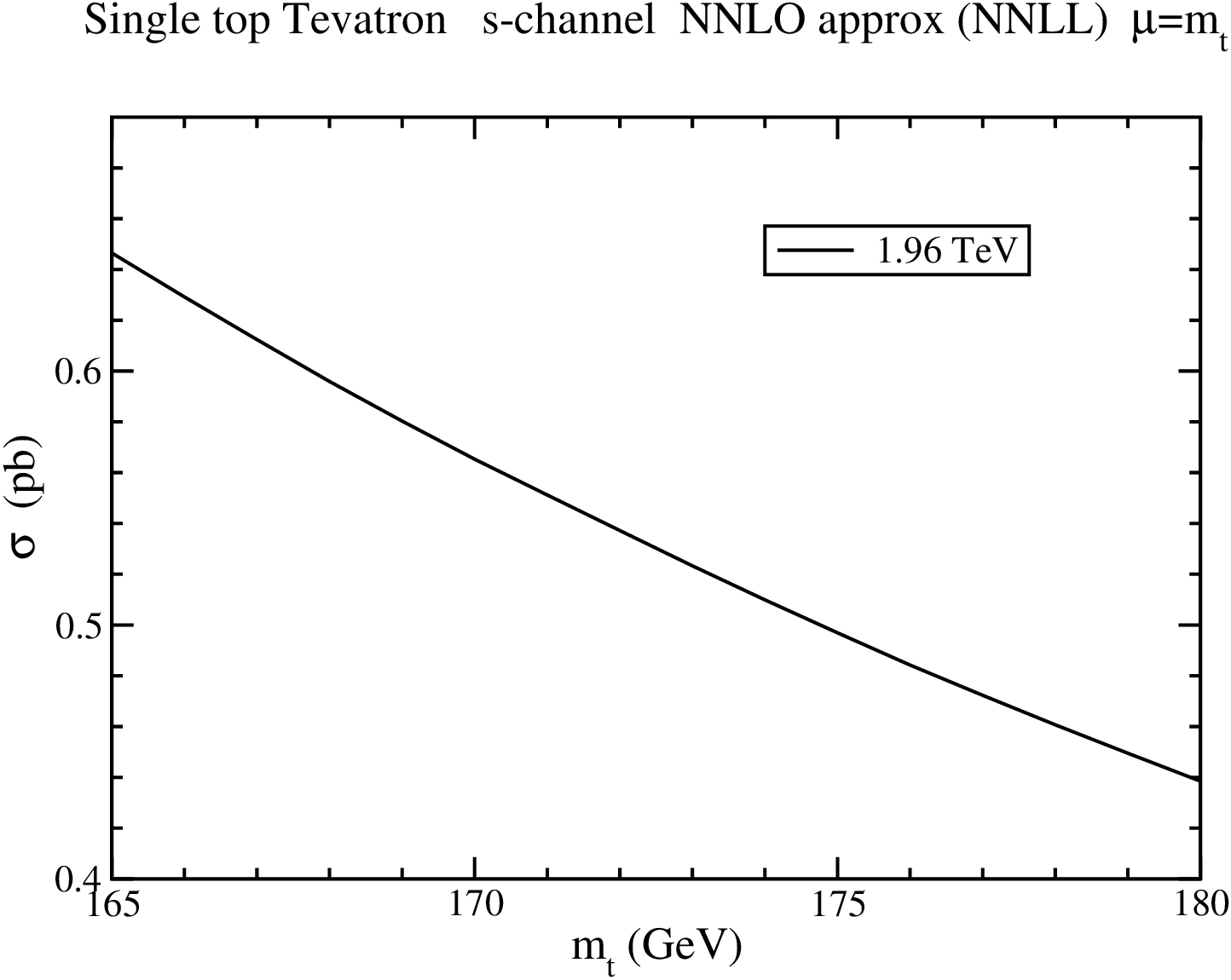}
  \caption{Single top $s$-channel cross section at the Tevatron.}
\end{center}
\end{figure}

For $s$-channel antitop production at the LHC we have 
$$
\sigma^{\rm NNLOapprox,\, antitop}_{s-{\rm channel}}(m_t=173\, 
{\rm GeV}, \, 7\, {\rm TeV})=1.42 \pm 0.01 {}^{+0.06}_{-0.07} \; {\rm pb}.
$$

Finally, we present results for associated $tW$ production,  
$bg \rightarrow tW^-$ \cite{NKtWH}. The cross section for this process
is very small at the Tevatron, but significant at the LHC.
We find that the NNLO approximate corrections increase the NLO cross section 
by $\sim 8$\% and 
$$
\sigma^{\rm NNLOapprox}_{tW}(m_t=173 \, {\rm GeV}, \, 7\, {\rm TeV})
=7.8 \pm 0.2 {}^{+0.5}_{-0.6} \; {\rm pb}.
$$
Figure 8 shows the $tW$ cross section at the LHC at both 7 TeV 
and 14 TeV energy. We note that the ${\bar t} W$ cross section is the same 
as that for $tW$ production.

\begin{figure}
\begin{center}
  \includegraphics[width=10cm]{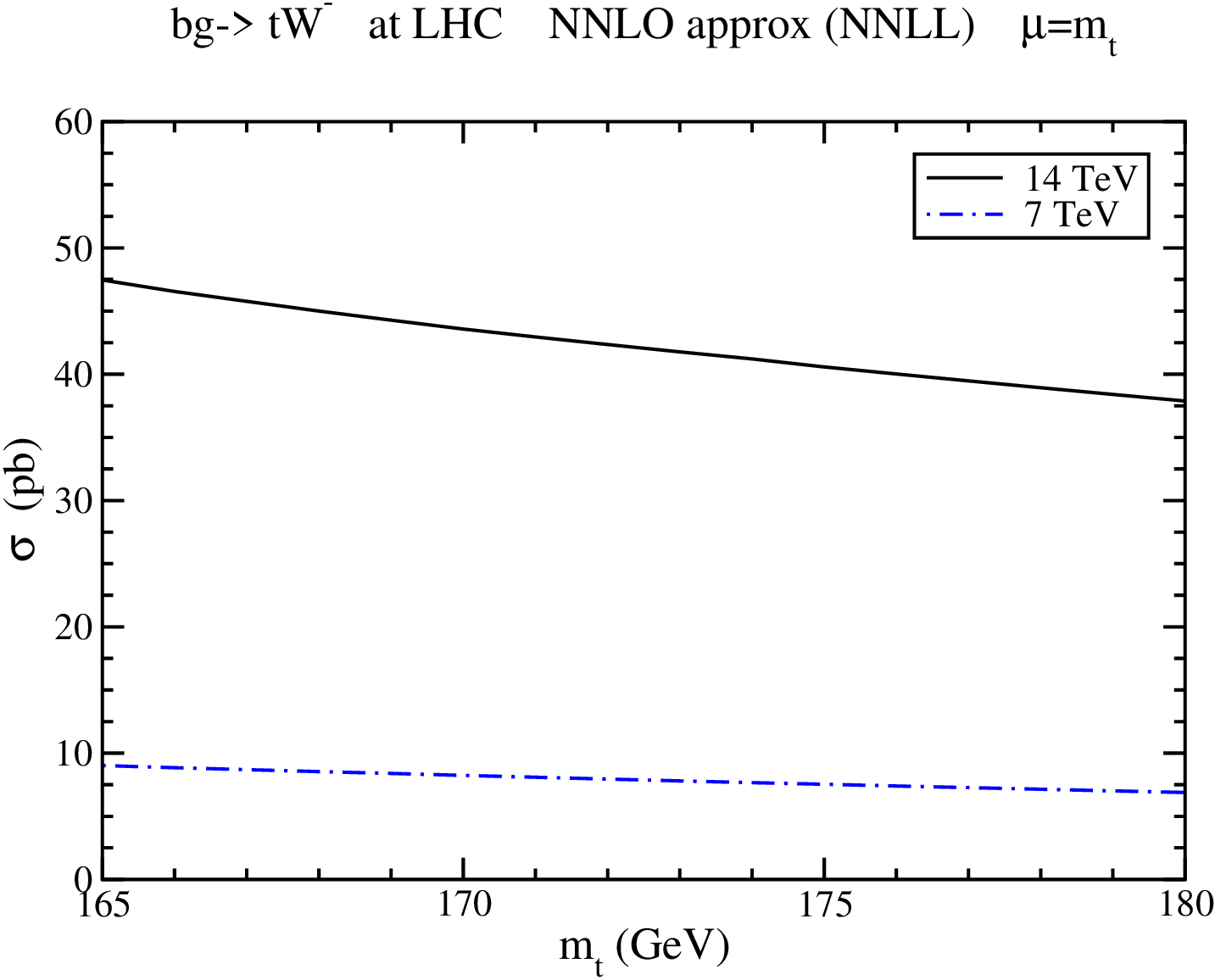}
  \caption{$tW$ cross section at the LHC.}
\end{center}
\end{figure}

A related process is associated charged Higgs production, 
$bg \rightarrow tH^-$, where the NNLO approximate corrections increase the 
NLO cross section by $\sim 15$ to $\sim 20$\% \cite{NKtWH}.

\vspace{3mm}

\begin{center}
{\large \bf ACKNOWLEDGMENTS}
\end{center}

This work was supported by the National Science Foundation under 
Grant No. PHY 0855421.

\end{document}